# Numerical simulation of Delft-jet-in-hot-coflow (DJHC) flames using the Eddy Dissipation Concept model for turbulence-chemistry interaction

Ashoke De, Ernst Oldenhof, Pratap Sathiah, Dirk Roekaerts

*Department of Multi-Scale Physics,*
*Delft University of Technology,*
*Lorentzweg 1, 2628 CJ Delft, The Netherlands.*
Tel.: +31-15-2782418
Fax.: +31-15-2781204
Email: D.J.E.M.Roekaerts@tudelft.nl

**Abstract**

In this paper we report results of a numerical investigation of turbulent natural gas combustion for jet in a coflow of lean combustion products in the Delft-Jet-in-Hot-Coflow (DJHC) burner which emulates MILD (Moderate and Intense Low Oxygen Dilution) combustion behavior. The focus is on assessing the performance of the Eddy Dissipation Concept (EDC) model in combination with two-equation turbulence models and chemical kinetic schemes for about 20 species (Correa mechanism and DRM19 mechanism) by comparing predictions with experimental measurements. We study two different flame conditions corresponding to two different oxygen levels (7.6% and 10.9% by mass) in the hot coflow, and for two jet Reynolds number (Re=4100 and Re=8800). The mean velocity and turbulent kinetic energy predicted by different turbulence models are in good agreement with data without exhibiting large differences among the model predictions. The realizable $k$-$\varepsilon$ model exhibits better performance in the prediction of entrainment. The EDC combustion model predicts too early ignition leading to a peak in the radial mean temperature profile at too low axial distance. However the model correctly predicts the experimentally observed decreasing trend of lift-off height with jet Reynolds number. A detailed analysis of the mean reaction rate of the EDC model is made and as possible cause for the deviations between model predictions and experiments a low turbulent Reynolds number effect is identified. Using modified EDC model constants prediction of too early ignition can be avoided. The results are weakly sensitive to the sub-model for laminar viscosity and laminar diffusion fluxes.

# 1. Introduction

MILD (Moderate and Intense Low Oxygen Dilution) combustion has been found to be one of the most promising techniques to improve the thermal efficiency of combustion systems (combustors, furnaces, burners etc) with lower pollutant emissions [1]. This technique is also known as HiTAC (High Temperature Air Combustion [2]) and FLOX (Flameless oxidation [3]), and all of these fall into the same category of 'clean combustion techniques'. In 'MILD' combustion, the inlet temperature of reactants remains higher than auto-ignition temperature of the mixture and, at the same time, the maximum temperature increase achieved during combustion usually remains lower than mixture auto-ignition temperature [1]. These conditions are reached by re-circulating the product gases into the incoming fresh air efficiently. The gas recirculation serves two purposes: (i) raise the reactant temperature (heat recovery) (ii) reduce the oxygen concentration (dilution). Because of the dilution with already cooled product gases the flame temperature is lower than usual (1100-1500 K), thereby reducing NOx emissions. This mode of combustion has several other characteristics such as hardly visible or audible flame, inherent flame stabilization, and semi-uniform temperature fields and smooth radiation flux yielding product quality in certain applications, i.e. petrochemical and metallurgical furnaces. In summary, MILD combustion technique is highly efficient and emits less pollutant compared to the conventional combustion techniques. Dally et al. [4] were the first to study flames of a jet-in-hot-coflow (JHC) burner with the objective of gaining better fundamental understanding of MILD combustion. We shall refer to the burner used in [4] as the Adelaide burner. The JHC burner emulates the MILD combustion conditions by realizing the mixed stream of combustion products and excess air with the help of a secondary burner, and primarily creates a MILD combustion zone in the mixing layer between a fuel jet and the mixed stream appearing as coflow surrounding the jet. An experimental database for CH4/H2 (50/50 by volume) fuel with 3%, 6% and 9% oxygen in the coflow was reported containing single point statistics from Raman-Rayleigh-LIF measurements [4]. Later in a slightly different burner design Medwell et al reported planar imaging of flame structures for several fuels [5, 6]. The Delft jet-in-hot-coflow (DJHC) burner was designed with the same objective of studying the fundamentals of MILD combustion. Results of experimental studies of JHC flames of natural gas were presented in Refs. [7, 8] and serve as validation data base for the present modeling study. The experimental database contains the results of high speed chemiluminescence imaging, velocity statistics from LDA measurements and temperature statistics from CARS measurements and qualitative OH-PLIF data [7, 8]. The studies of the DJHC burner are complementary to ongoing experimental and numerical studies in Delft of combustion in a 300 kW furnace equipped with several FLOX burners [9]. The combined aim of both studies is to provide insights useful for modeling industrial furnaces operating in flameless mode.

In general, low oxygen concentration and moderate temperature levels lead to slower reaction rates. Hence, for flames in the MILD regime finite rate chemistry is more important than usually the case for conventional diffusion flames. At the same time mixing processes remain essential: on one hand the mixing process of products with either fuel or air, on the other hand the mixing of the diluted fuel and oxidizer streams. In the JHC burner the relevant mixing process occurs in-between the undiluted fuel jet and the hot diluted coflow. In the presence of concurrent turbulent mixing and chemical reaction between the mixed streams, turbulence-chemistry interaction appears as an important modeling problem. Several approaches to turbulence-chemistry interaction modeling have already been applied to MILD combustion in the literature. Christo and Dally modeled the JHC burner experiments of Ref. [4] using steady flamelet model, Eddy Dissipation Concept (EDC) model and transported Probability Density Function (PDF) model [10, 11]. These validation studies considered three levels of oxygen concentration in the coflow studied experimentally: 3%, 6% and 9% and concluded that the EDC model produced better results than the flamelet model. The transported PDF predictions were of comparable quality as the EDC model predictions [11], but found sensitive to the level of velocity fluctuations which could not be validated independently. Moreover, they also reported that differential diffusion played an important role in their flames due to composition of the fuel ($CH_4/H_2$ in ratio 50/50 by volume). Several other simulation studies of the experiments of Ref. [4] have been published. Kim et al. [12] simulated the same burner using conditional moment closure (CMC) model and their results showed that the effects of oxygen concentration on the flame structure and NO formation were well predicted; however their simulations did not include differential diffusion effects. Frassoldati et al. [13] numerically investigated the same burner to study the effect of inlet turbulence and applied a detailed NOx post-processor. Mardani et al [14] studied the sensitivity of the predictions to different models for differential diffusion in the context of the EDC model. The JHC flames emulating MILD combustion have lower oxygen concentration in the coflow than a related set of lifted flames in hot coflow studied in the literature. But also in those related flames extensive modeling studies have been reported, starting with a study using the EDC model by Myhrvold et al. [15]. Recently Ihme and See [16] applied large eddy simulation with a steady flamelet / progress variable model to the experiments of Ref. [4] and found good agreement provided the burner was described using a three stream problem described by two mixture fractions. The first mixture fraction is associated with the mixing of fuel and coflow. The second mixture fraction described both the entrainment of surrounding air and non-homogeneity of the radial oxygen profile in the coflow region. This illustrates the complexity of applying conventional two-stream flamelet models to JHC burner emulating MILD combustion behavior.

MILD combustion has also been studied in laboratory scale furnaces [17-27] and semi-industrial scale furnaces. For example, Orsino and Weber [24] investigated a semi-industrial scale burner using three different turbulent-chemistry interaction models (Eddy Break up Model (EBU), Eddy Dissipation Concept (EDC) and presumed PDF with equilibrium chemistry). They reported that all of these three models were capable of predicting the main flow characteristics properly, such as temperature fields, high radiative flux and low NOx and CO. Mancini et al. [25] had found that prediction of the entrainment by jets is a crucial element for successful modeling of MILD combustion burners. Kim et al. [26] studied the effects of different global reaction mechanisms for modeling MILD combustion of natural gas. More recently, the experimental and numerical investigation by Danon et al. [27] reported an industrial scale FLOX$^{TM}$ burner firing at low calorific value gases. Their analysis showed that Eddy Dissipation Concept (EDC) combustion model in conjunction with a skeletal chemistry mechanism reasonably well predicted the axial profiles of mean temperature and species concentration in the flue gas while compared to experimental values. The EDC model has been applied in modeling of gasification by Rehm et al [28]. They reported that for this system with relatively slow reactions, model adjustments were necessary.

From this short literature review it can be concluded that the EDC model has been widely applied to MILD combustion systems, including JHC burners, and performs reasonably well compared to other standard combustion models. Nevertheless a detailed analysis of the quality of its predictions is missing. Therefore it is worthwhile to use the database of the DJHC burner experiments [7, 8] to make such analysis. The objective of the study reported here is to assess the performance of the EDC models as turbulence-chemistry interaction model in combination with different turbulence models, and chemical mechanisms in predicting JHC flames by comparing predictions with experimental data. A detailed dataset obtained from measurements (mean velocity, Reynolds stresses, mean temperature and its standard deviation) is used as boundary conditions for the numerical models and the predictions of both velocity statistics and temperature statistics are used as validation criterion. The availability of velocity measurements has a decisive advantage. Indeed, in modeling studies of the experiments of [4], a strong sensitivity of predictions on inlet boundary conditions for the turbulent kinetic energy was observed [13]. Since no experimental data on this quantity were available it could be used as a variable parameter in the model. The DJHC burner experiments offer an opportunity to further investigate the performance of the EDC model, but not as a simple repetition, but as a more thorough test. The measured velocity field statistics fix the boundary conditions and provide quantities to be used for model validation. The article is structured as follows. Firstly, a description of burner and experiments is given. Next, the turbulent combustion models are described. Then the results of the simulations are discussed and finally conclusions are presented.

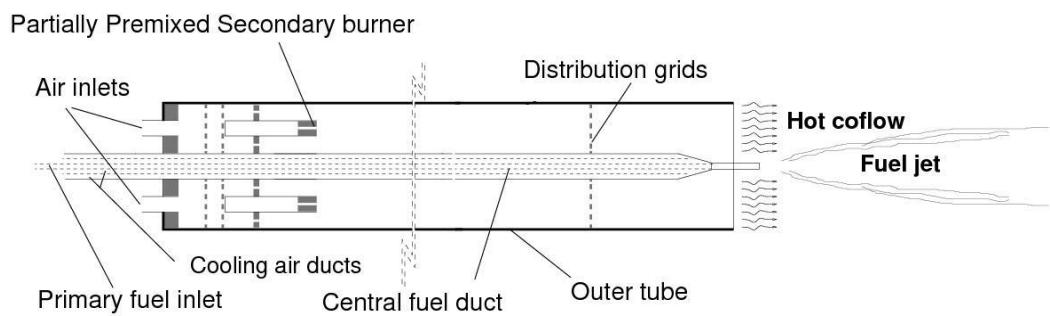

Figure 1. Schematic of DJHC burner

## 2. Description of the burner and experiments

### 2.1 Burner design

The design of the Delft jet-in-hot coflow (DJHC) burner is based on that of the Adelaide jet-in-hot coflow burner [4]. In both cases a fuel jet enters in a coflow with relatively low oxygen concentration and sufficiently high temperature to have a mixture above auto-ignition temperature after mixing. A sketch of the design of the DJHC burner is shown in Fig. 1. The primary burner consists of a central primary fuel jet with 4.5 mm inner diameter. The coflow is generated by the secondary burner in an annulus of diameter 82.8 mm. The secondary burner combines premixed and non-premixed combustion. It consists of a ring of premixed flames with additional air injected at both sides of the ring. The partial premixing is found to give stable combustion. The central fuel pipe is cooled by a continuous flow of air through the concentric cooling air ducts, thus preventing excessive heating of the main fuel jet. More detailed description can be found in in references [7, 8] and for illustration in the Appendix some additional photographs are shown of burner and flames.

It is important to independently control oxygen content and temperature of the coflow in order to mimic the conditions in a real furnace, where flue gas entrained aerodynamically into the near burner zone after losing heat to the surroundings. To reach this goal, in the experiments with the Adelaide JHC burner [4] $N_2$ was added as an independent third stream. In the DJHC the coflow is created in the secondary burner on the basis of only natural gas and air as reactants and the coflow temperature is controlled by influencing the radiative and convective heat losses from the burner pipe using one or more perforated screens positioned in the coflow annulus. In this report we describe experiments using only one such distribution grid located 11cm upstream of the burner exit plane (here defined as end of wall of the outer wall). It is observed that the outer wall of the burner tube radiates most strongly at the height of the distribution grid. Another function of the distribution grids is to mix the product flow in circumferential and radial direction. Experimentally it is found that the coflow indeed is almost perfectly axi-symmetric, but there remains a radial dependence in the temperature and the composition, described in more detail below, and this has to be taken into account in the modeling.

Another difference between the Adelaide and the Delft burner is found in the possibilities for adding seeding particles to the coflow. By avoiding small sized obstructions such as screens or porous elements, the DJHC allows for the addition of seeding particles in the burner to act as tracers in for instance LDA or PIV measurements. The major flow of air into the coflow passes through the air inlets at the bottom and it is this flow that carries the seeding particles for LDA measurements.

| Case | Jet $\tilde{u}_j$ (m/s) | Jet $Re_d$ | Coflow Sec. fuel (nl/min) | Coflow Total air (nl/min) | $T_{max;co}$ (K) | $Y_{O2;co}$ (%) |
|---|---|---|---|---|---|---|
| DJHC-I_S | 34.0 | 4100 | 16.1 | 224 | 1540 | 7.6 |
| DJHC-X_S | 32.9 | 4600 | 14.2 | 239 | 1395 | 10.9 |
| DJHC-I_S | 56.0 | 8800 | 16.1 | 224 | 1540 | 7.6 |

Table 1. Characteristics of jet and coflow in the DJHC burner experiments. $T_{max;co}$ is the maximum mean coflow temperature, $Y_{O2;co}$ is the calculated mass flux weighted mean mass fraction in the coflow.

**2.2 Experimental data base**

A series of experiments in the DJHC burner has been made, at various coflow temperatures, coflow oxygen level, and for different fuel jet Reynolds number [7, 8]. In addition to Dutch natural gas also pure methane and methane/ethane mixture has been used as fuel. The modeling study reported here concerns natural gas combustion, with two different types of coflow as reported in Table 1. The DJHC-X_S flame has higher oxygen content in the coflow and lower temperature than the DJHC-I_S flame, leading to different ignition phenomena [7, 8]. Observation using high speed camera has revealed that the main flame is stabilized by randomly formed auto-ignition spots that grow while being convected downstream by the mean flow [7].

Laser diagnostic measurements have been made along radial profiles at different axial locations, X=0.003, 0.015, 0.03, 0.060, 0.09, 0.12 and 0.15 m from the jet exit. In addition an axial profile is also measured at the centerline of the jet. From multiple instantaneous measurements of velocity using Laser Doppler Anemometry (LDA), the Favre averaged velocity, $\tilde{u}_j$ and the Reynolds stress, $u_i'' u_j''$ are obtained. From multiple instantaneous measurements of temperature using Coherent Anti-stokes Raman Spectroscopy (CARS), the mean temperature and its variance are obtained. The radial profile of oxygen concentration in the coflow was obtained using probe measurements.

## 3. Model description

### 3.1 Introduction

In MILD combustion ignition and flame stabilization take place after mixing of the fuel and diluted oxidizer streams and this mixing process most often occurs in turbulent conditions due to the use of high momentum jets. In the experiment considered here the oxidizer stream is diluted with products and it is the scalar mixing of the fuel and oxidizer streams that has to be described accurately. This is only possible if the underlying turbulence velocity field is modeled accurately. Due to the dilution of the coflow, the stoichiometric mass ratio of fuel and co-flow is rather low and the scalar mixing layer is at larger radial distance than the momentum shear layer, in a region with relatively low level of turbulence. Depending on the composition and temperature of the coflow the ignition of fresh mixture could be by auto-ignition or by flame propagation. In both cases a sufficiently detailed chemical mechanism has to be used. Finally a model for the closure of the mean chemical source term has to be chosen. For the current study a combination of two-equation turbulence models, reduced chemical mechanisms for 16 to 21 species and the Eddy Dissipation Concept (EDC) model for closure of the mean chemical source term are chosen. In addition a modeled equation for the variance of temperature is solved.

### 3.2 Turbulence and chemistry models

To model turbulence, three different models have been used: the standard k-$\varepsilon$ model (SKE), the realizable k-$\varepsilon$ model (RKE) [29] and the renormalization group k-$\varepsilon$ model (RNG) [30]. In all three models the Reynolds stress is closed using mean velocity gradients employing Boussinesq hypothesis as follows:

$$-\overline{\rho u_i u_j} = \mu_t \left( \frac{\partial u_i}{\partial x_j} + \frac{\partial u_j}{\partial x_i} \right) - \frac{2}{3} \left( \overline{\rho} k + \mu_t \frac{\partial u_k}{\partial x_k} \right) \delta_{ij} \tag{1}$$

where the turbulent eddy viscosity is calculated by different relationships in the three models. The simulations are carried out using different turbulence models with the aim of selecting the model giving best prediction for velocity statistics in order to get the detailed analysis of thermo-chemical fields.

Also different chemical mechanisms are used in the simulations. Presented below are results obtained using the mechanism proposed by Correa in Ref. [31], having 16 species and 41 reaction steps and the mechanism called DRM19, documented in Ref. [32]. The DRM19 mechanism is obtained by reduction of GRI 1.2 [33] and has 21 species and 84 reactions. A brief literature review of its reported performance is given by Mardani et al [14].

### 3.3 Turbulence-chemistry interaction model

#### 3.3.1 Turbulence scales and the Eddy Dissipation Concept (EDC) model

The Eddy Dissipation Concept Model, developed by B. Magnussen et al. [34, 35], has the advantage of incorporating the influence of finite rate kinetics at computational cost which is quite moderate compared to more advanced models as the transported PDF method. This advantage comes at the cost of a less accurate description of turbulent temperature fluctuations. It is important to look at the EDC model description in detail in order to be able to correctly interpret the results. So in the next subsections we first present details of the EDC model and study the dependence on turbulent Reynolds number and the sensitivity to values of model constants. In the EDC model the influence of turbulent fluctuations on mean chemical reaction rate is taken into account by reference to the phenomenological description of turbulence in terms of the turbulent energy cascade. Key variables in the description using the turbulent energy cascade are the energy dissipation rate ε and the characteristics of the viscous scale of the flow (Kolmogorov scale), only depending on ε and the kinematic viscosity ν. Let the length, time and velocity scales of the energy containing range of the spectrum be denoted by $l_t, T_t$ and $u'$. Then the turbulence Reynolds number is defined by

$$\text{Re}_t = \frac{u' l_t}{\nu} \tag{2}$$

The velocity and length scales of the energy containing range are determined using a turbulence model.

In the frame of the k-ε models $T_t = \dfrac{k}{\varepsilon}$, $u' = \sqrt{k}$, $l_t = \dfrac{k^{3/2}}{\varepsilon}$, $\mathrm{Re}_t = \dfrac{k^2}{\nu\varepsilon}$ with both k and ε calculated from model transport equations. The Kolmogorov scales for length, time and velocity η, $\tau_K$ and $u_K$ are related to the scales of the energy containing range via the turbulence Reynolds number:

$$\mathrm{Re}_t = \left(\dfrac{T_t}{\tau_K}\right)^2 = \left(\dfrac{l_t}{\eta}\right)^{4/3} = \left(\dfrac{u'}{u_K}\right)^4 \qquad (3)$$

The quantity $\mathrm{Re}_t$ is in fact proportional to the ratio of turbulent viscosity to laminar viscosity.

### 3.3.2 EDC model equations

In the following we shall first present the model, then discuss sensitivity to model constants and finally discuss an interpretation of the model as manifold method.

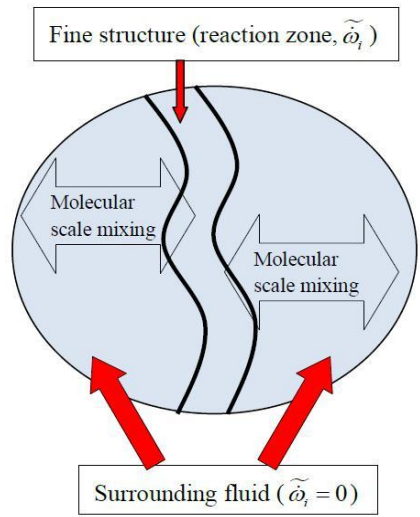

Figure 2. Schematic of computation cell based on EDC model

**Model for the mean chemical source term**

Recognizing the fact that chemical reaction takes place within a thin confined reaction zone which is typically smaller than the size of the computational grid, in the arguments underlying the Gran and Magnussen EDC model the computational cell is conceptually divided into two sub-zones: the reacting "fine structure" and "the surrounding fluid" as shown in Figure 2. All the homogeneous chemical reactions are assumed to occur in the fine structures that are locally treated as adiabatic, isobaric, Perfectly Stirred Reactors (PSR) transferring mass and energy only to the surrounding fluid [34] where only turbulent mixing takes place (without chemical reaction), thereby transporting the surrounding reactant and product gases to and from the fine structure. The size of the fine structure $\xi$ and also the mean residence time of the fluid within the fine structure $\tau$ are determined and described below (Eqs 8 & 9). Assuming that chemical reactions take place only in the fine structure, the reaction rates of all species are calculated from a mass balance of the fine structure reactor and the net mean species reaction rate for the transport equation is given by [34]:

$$\dot{\omega}_i = \frac{\overline{\rho}\xi^2 \lambda}{\tau}(Y_i^* - Y_i^o) \tag{4}$$

where $Y_i^*$ and $Y_i^o$ represent the mass fractions of species, $i$, in reacting (fine structure) and non-reacting (surrounding fluid) part, respectively. The mean residence time, $\tau$ is the inverse of the specific mass exchange rate between fine structures and surroundings m*: $\tau = \frac{1}{m^*}$. The factor, $\lambda$, is the fraction of fine structures where reaction takes place. The mean mass fraction $Y_i$ can be obtained from the linear combination of properties in the fine structures and the surrounding fluid as follows ([34], Eq. 6):

$$Y_i = \xi^3 \lambda Y_i^* + (1 - \xi^3 \lambda) Y_i^o \tag{5}$$

Sensitivity analysis performed in Ref. [30] has led to the recommendation that when using EDC with detailed chemistry the choice $\lambda = 1$ is best. Putting this value in Eq. (5) the value of $Y_i^o$ is obtained as:

$$Y_i^o = \frac{Y_i - \xi^3 Y_i^*}{(1 - \xi^3)} \tag{6}$$

Rearranging the Eq. (4) and Eq. (6), the expression for mean chemical source term in the transport equation for the mean mass fraction $\tilde{Y}_i$ is obtained as

$$\dot{\omega}_i = \frac{\overline{\rho}\xi^2}{\tau(1 - \xi^3)}(Y_i^* - Y_i) \tag{7}$$

In many articles in the literature use has been made of a variant of the EDC model, where the combustion in the fine structure is assumed to occur in a constant pressure reactor, with initial conditions taken as the current species and temperature in the cell, instead of a PSR. In this approach reactions proceed over the time scale, τ, governed by Arrhenius rates, and are integrated numerically in time. This means use is made of "Plug Flow Reactor" (PFR), i.e. it is integrating reaction kinetics in time, rather than "Perfectly Stirred Reactor (PSR)" (involving mixing of final products with fresh reactants in one volume). Only in the case of slight exothermicity and very small conversion the two reactor models PSR and PFR give the same output. In particular the studies using the CFD code Fluent [29], including the current study, use the PFR.

The size of the fine structure ξ and the mean residence time τ in EDC model is calculated as

$$\xi = C_\xi \left(\frac{\nu \varepsilon}{k^2}\right)^{1/4} = C_\xi \left(\frac{l_t}{\eta}\right)^{-1/3} = C_\xi \left(\text{Re}_t\right)^{-1/4} \tag{8}$$

and

$$\tau = C_\tau \left(\frac{\nu}{\varepsilon}\right)^{1/2} = C_\tau \tau_K = C_\tau \text{Re}_t^{-1/2} T_t = C_\tau \text{Re}_t^{-1/2} \frac{k}{\varepsilon} \tag{9}$$

where the model constants have the following default values: $C_\xi = 2.1377$, $C_\tau = 0.4082$. These values follow from the values of related model constants $C_{D1}$ and $C_{D2}$ mentioned in Refs. [34, 35].

The variable $Y_i^*$ is the species mass fraction reached from the current value of $\tilde{Y}_i$ by action of the applied chemical reaction mechanism over a time scale $\tau$. So in the EDC model the mean chemical state evolves via a linear relaxation process, typical for mixing, towards a reacted state which would be reached by a nonlinear reaction process after a time scale $\tau$. However from Eq. (7) it can be seen that the characteristic time scale of the linear relaxation process is of the order of the time scale of the energy containing scales of turbulence, which is larger than $\tau$:

$$\frac{1}{\tau_{mix}} = \frac{\xi^2}{(1-\xi^3)} \frac{1}{\tau} = \frac{1}{(1-\xi^3)} \left(\frac{C_\xi^2}{C_\tau} \frac{1}{T_t}\right) \tag{10}$$

The factor

$$\frac{1}{(1-\xi^3)} = \left(1 - C_\xi^3 \frac{\eta}{l_t}\right)^{-1} \qquad (11)$$

represents a correction for the fact that the fluid receiving mass by mixing with the reacted material does occupy only part of the volume of the cell [16]. The fact that the volume ratio is proportional to a length scale ratio is in agreement with a view that reaction zones are two-dimensional sheets.

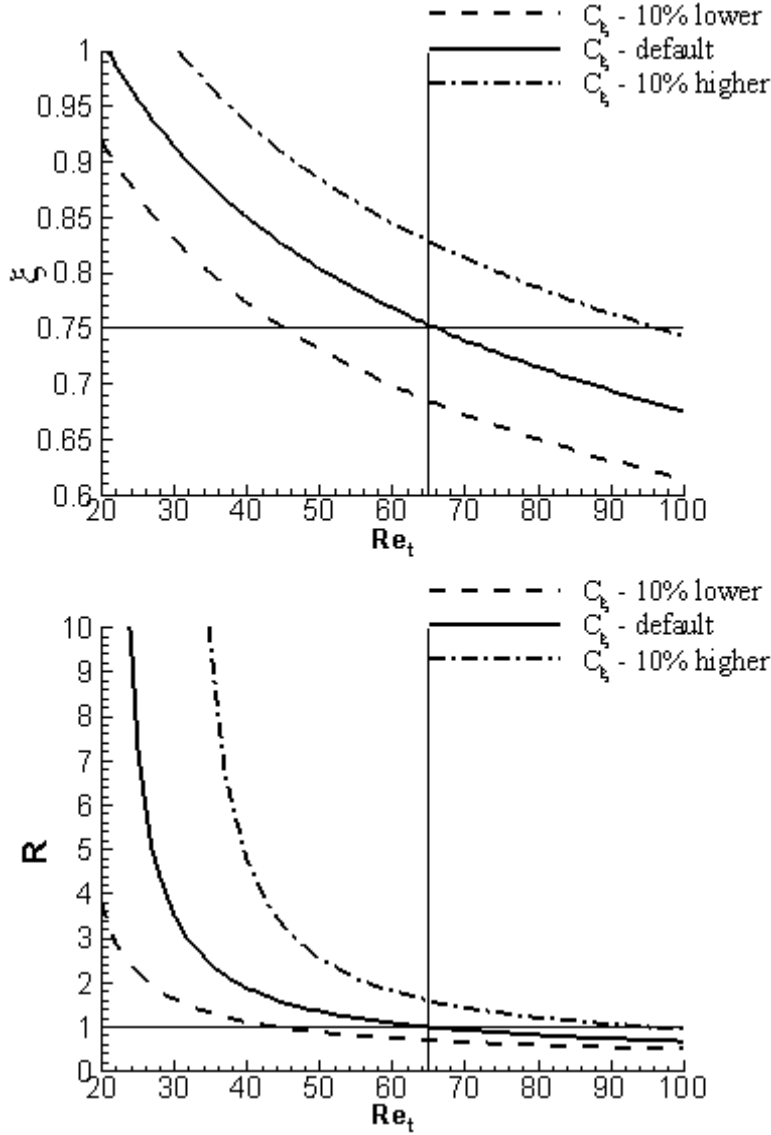

Figure 3: Dependence of 'size of fine structure' parameter $\xi$ and time scale ratio (R) on turbulent Reynolds number for three different values of the model constant $C_\xi$

The evolution of $Y_i^*$ after reacting over time $\tau$, depends on the chemical mechanism used in computation. To reduce the computational time for integrating the Arrhenius reaction rates in these fine scales, the In Situ Adaptive Tabulation (ISAT) method of Pope [36] is used. In the EDC calculations reported here the ISAT error tolerance is set to $10^{-5}$. For the quantities investigated, there was no difference found in the results using tolerance factor $10^{-3}$ instead of $10^{-5}$ and the computational effort decreases 8-10 times if the tolerance is set as $10^{-3}$ instead of $10^{-5}$. For the current study however the computational effort is not excessive and a tolerance $10^{-5}$ is maintained.

**Model analysis and sensitivity to model constants**

Note that in the EDC model the evaluation of kinetic rates, in particular also the evaluation of the temperature dependent Arrhenius factors are done using the mean concentrations and the mean temperature. The influence of temperature fluctuations on mean reaction rates is not taken into account using information on the statistics of the temperature field, but only using information on the statistics of the velocity field as represented in the turbulence model. The fact that in MILD combustion temperature fluctuations are significantly lower than in flames without dilution seems a necessary condition to make this a successful modeling strategy.

It is of interest to check the sensitivity of the predicted reaction rate to the model constants $C_\tau$ and $C_\xi$. The model constant $C_\tau$ appears as a multiplicative factor in the time scale $\tau$ and hence also in the reaction rate term. The sensitivity of the reaction rate to its value mainly is through its influence on $Y_i^*$ which is depending on the detailed chemistry. The dependence on the model constant $C_\xi$ is more intricate. Figure 3 shows the sensitivity of $\xi$ and of the time scale ratio

$$R = \frac{\tau}{\tau_{mix}} = \frac{\xi^2}{\left(1-\xi^3\right)} \qquad (12)$$

For consistency $\xi$ and also R must be lower than one. The second condition (R<1) is more severe than the first one ($\xi$<1) and, for default values of the model constant $C_\xi$, it limits the applicability of the model expression (7) to values $\mathrm{Re}_t > 65$. For R=1, and using the PFR model, the expression (7) reduces to

$$\dot{\omega}_i = \bar{\rho}\frac{(Y_i^{PFR}(t+\tau)-Y_i(t))}{\tau} \qquad (13)$$

Here $Y_i^{PFR}(t+\tau)$ is the value of the plug flow reactor at $t+\tau$ starting from initial condition $Y_i(t)$. Expression (13) can be interpreted as a first order discretization of the laminar kinetics, an observation that will be used in the discussion of the model performance below.

**Interpretation as a manifold method**

The EDC model is formulated on the basis of a qualitative picture which may or may not be valid in the case of MILD combustion. It is of interest to re-interpret the EDC model equations from the point of view of evolution in the chemical state space, similar to what is done in reduced chemistry formulations using low dimensional manifolds in thermochemical phase space.

Let the mean species concentration in a computational cell at time $t_0$ be given by $Y_i(\vec{x},t_0)$. According the detailed chemical mechanism it would evolve after the time $\tau$, of the order of the Kolmogorov time scale, to a new value $Y_i(\vec{x},t_0+\tau)$. This evolution from $Y_i(\vec{x},t_0)$ to $Y_i(\vec{x},t_0+\tau)$ in general will be a curved trajectory in the thermochemical phase space. The EDC model replaces this evolution according the detailed chemical kinetics by a evolution along a straight line in phase space connecting $Y_i(\vec{x},t_0)$ to $Y_i(\vec{x},t_0+\tau)$. In this way transients with a chemical time scale shorter than the Kolmogorov time scale do not show up in the evolution of the chemical states of the computational cells. The motion in phase space along the straight line in phase space is not at a uniform rate, but in the form of an exponential. The rate of the motion along the straight line depends on the time scale of the energy containing eddies $\tau_{mix}$ and as shown above at low turbulent Reynolds number can be sensitive to the model constant $C_\xi$.

Assuming that the CFD calculation numerically proceeds with a finite time step $\Delta t$ smaller than the time scale $\tau$ and assuming that the 'target state' $Y_i(\vec{x},t_0+\tau)$ is updated every time step, the manifold of states reached during evolution according the EDC model consists of trajectories in phase space with tangent vector in the direction of the line connecting current state and moving 'target state' $Y_i(\vec{x},t_0+\tau)$. The actually followed trajectory in state space then is not linear but curved.

The most important consequence of the followed procedure is that the chemical effects which take less than a Kolmogorov time scale are washed out. Indeed, the current state relaxes linearly to a target state. The target state is the state the system would reach after one Kolmogorov time scale. The relaxation time differs from the Kolmogorov time scale, but is in fact a slower time scale, so that by the linear relaxation the system does not reach the target state. Some chemical phenomena (either production or destruction) might be homogeneous in time over a Kolmogorov time scale. But minor species, typically affected by fast chemical phenomena, could be formed and again destroyed just in one Kolmogorov time scale. In that case the local maximum is overlooked by the linear relaxation. Effectively the algorithm then describes the evolution in a sub-manifold of the phase space, showing resemblance to the evolution on trajectory generated low-dimensional manifolds.

### 3.3.3. Extension of EDC model with equation for temperature variance

The reaction rates in the EDC model are evaluated using the mean temperature as obtained by solving the energy equation [29]. Because measurements of temperature variance are available it is of interest to extend the EDC model with an equation for the temperature variance. Assuming weak dependence of mean specific heat on composition and temperature, the following modeled equation for the variance can be proposed

$$\frac{\partial(\overline{\rho}T''^2)}{\partial t} + \frac{\partial(\overline{\rho}u_i T''^2)}{\partial x_i} = \frac{\partial}{\partial x_i}\left[(\mu + \frac{\mu_t}{\sigma_t})\frac{\partial T''^2}{\partial x_i}\right] + C_p \mu_t \left(\frac{\partial T}{\partial x_i}\right)^2$$

$$-C_d \overline{\rho}T''^2 + 2\overline{\rho}T''^2 \dot{\omega}_T \qquad (14)$$

where $C_p = 2.86$, $C_d = 2.0$ and $\sigma_t = 0.85$ are model constants. In the derivation of this equation a gradient diffusion assumption has been used to model the turbulent scalar flux and a simple relaxation assumption to model the dissipation rate of temperature variance. The last term on the right hand side of Eq. 14 represents the production of temperature fluctuations due to heat release by chemical reaction, $\dot{\omega}_T$. For simplicity, the contribution of this term is assumed to be negligible. The variance model has no influence on the mean reaction rate of the EDC model can be considered a post-processing model. In the comparison between modeling and measurements below we assume that the experimental mean and standard deviation from CARS can be directly compared to computed Favre averages.

## 3.4 Setup of the CFD calculation

### 3.4.1 Computational domain and grid

Because of the symmetry of the burner a 2D axisymmetric grid is used. The computational domain starts 3 mm downstream of the jet exit and extends up to 225 mm in the axial direction. In the radial direction, the grid extends up to radial distance 80 mm to take into account the effect of entrainment of ambient cold air. The grid consists of 180x125 cells, in axial and radial direction, respectively, stretched both in axial and radial directions. Also two finer meshes, having 270x187 and 360x250 cells, have been used in a grid independence study.

### 3.4.2 Boundary conditions

The boundary of the computational domain at the highest axial distance is set to outflow. Symmetry is imposed at the lateral boundary and inflow at the burner exit. At the inlet boundary for both jet and coflow, the values of physical quantities are set to be equal to the measured values at the first measured radial profile (x=3 mm) [7,8]. Mean velocity, mean temperature and variance of temperature are present in the experimental database. (The measured data are interpolated on the numerical mesh). The turbulent kinetic energy, k, at the inlet boundary, is calculated from the measured axial and radial normal components of the Reynolds stress, and assuming the azimuthal component to be equal to radial one i.e., $w'^2 = v'^2$ ($k = (u'^2 + v'^2 + w'^2)$). To construct a value of turbulence dissipation rate ε at the inlet, it is assumed that the turbulence production is equal to dissipation, and ε is calculated using Eq. 15 from the measured quantities:

$$\varepsilon = -u'v'\frac{\partial u}{\partial r} \tag{15}$$

The fuel composition is specified to be one of the two mixtures approximating Dutch natural gas specified in Table 2. Fuel I consists of methane and nitrogen and fuel II consists of methane, ethane and nitrogen. Fuel II can only be used when the chemical mechanism applies to ethane, i.e. with the DRM19 mechanism, but not with the Correa mechanism. Both mixtures are defined to have the same calorific value as Dutch natural gas.

| Mole % | Fuel I | Fuel II | Dutch natural gas |
|---|---|---|---|
| $N_2$ | 15.0 | 15.0 | 14.4 |
| $CH_4$ | 85.0 | 81.0 | 81.3 |
| $C_2H_6$ | - | 4.0 | 3.7 |
| Rest | - | - | 0.6 |

Table 2. Fuel compositions used in this study

The inlet boundary condition for mean oxygen mole fraction is taken from the experimental values measured at x=3 mm. Other species concentrations are calculated using an equilibrium assumption. This can be done as follows. The coflow is considered to be a stream of non-adiabatic equilibrium products of combustion of Dutch natural gas. The known overall fuel/air ratio determines the overall equivalence ratio or overall mixture fraction. The local mean oxygen mole fraction determines the mean mixture fraction as function of radial distance. The difference between the measured mean temperature and the adiabatic equilibrium temperature determines the enthalpy deficit as function of radial distance. Taking into account the enthalpy deficit the corresponding equilibrium composition is calculated using the TU Delft FLAME code [37]. Finally we remark that the radial extent of the computational domain is larger than the width of the coflow. The temperature at the inlet drops from the outermost measured value in the coflow to room temperature over a distance of 5 mm.

### 3.4.3 Computational algorithm

The flow simulations are done using Fluent-12.0.16 [29]. A steady solution of the mean transport equations is computed. In the solution of the mean transport equations and the turbulence model, a SIMPLE algorithm is used for pressure-velocity coupling. The second order discretization scheme is consistently used for all the terms.

## 4. Results

In this section, we present simulations results and compare with the measurements mainly for flame DJHC-I_S at Re=4100 and Re=8800. (Predictions for DJHC-X_S at Re=4600 are added at the end of section 4.5) We first examine grid independence. Then we discuss predictions of the mean axial velocity, turbulent kinetic energy, and turbulent shear stress. Finally we analyze the predictions of mean temperature, temperature variance and lift-off height. The sensitivity of the predictions to the choice of turbulence model, chemical kinetic mechanism and values of model constants in the EDC model is studied.

### 4.1 Grid independence

In order to investigate the grid independence of the predictions for the case with Re=4100 and with low $O_2$ concentration, DJHC_I-S, Re=4100, as given in Table 1, we made a simulation with a fixed model on three different grids: one consists of 180x125 (axial x radial) cells; the other two finer grids contain 270x187 and 360x350 cells, respectively. The simulations are done using standard $k$-$\varepsilon$ model and EDC model with Correa chemical mechanism.

Figure 4 shows the predicted results obtained from three different grids for mean axial velocity ($U_x$) and turbulent kinetic energy ($k$) at two different axial locations and along the center line as well. As observed, the predictions show proper trends of jet spreading and evolution of turbulent kinetic energy, and moreover the simulations using different grids are in good agreement with each other. However, it should be noted that the spreading rate of the jet is slightly over-predicted, as visible in the radial profiles of axial velocity and turbulent kinetic energy. Below it is shown that the realizable k-$\varepsilon$ model is performing better in this respect. Since the simulations using different grids give the same results, the coarser grid with 180x125 cells is chosen here for the rest of the reacting flow calculations.

## 4.2 Prediction of velocity statistics

In this section, the predicted mean velocity, turbulent kinetic energy and Reynolds shear stress are compared to measurements. We have compared the predictions of SKE, RKE, RNG and SKE with modified model constants in combination with the EDC model with the Correa chemical mechanism [26]. It is found that the differences between the models are larger at Re=8800 then at Re=4100. The RNG model does not give satisfactory results.

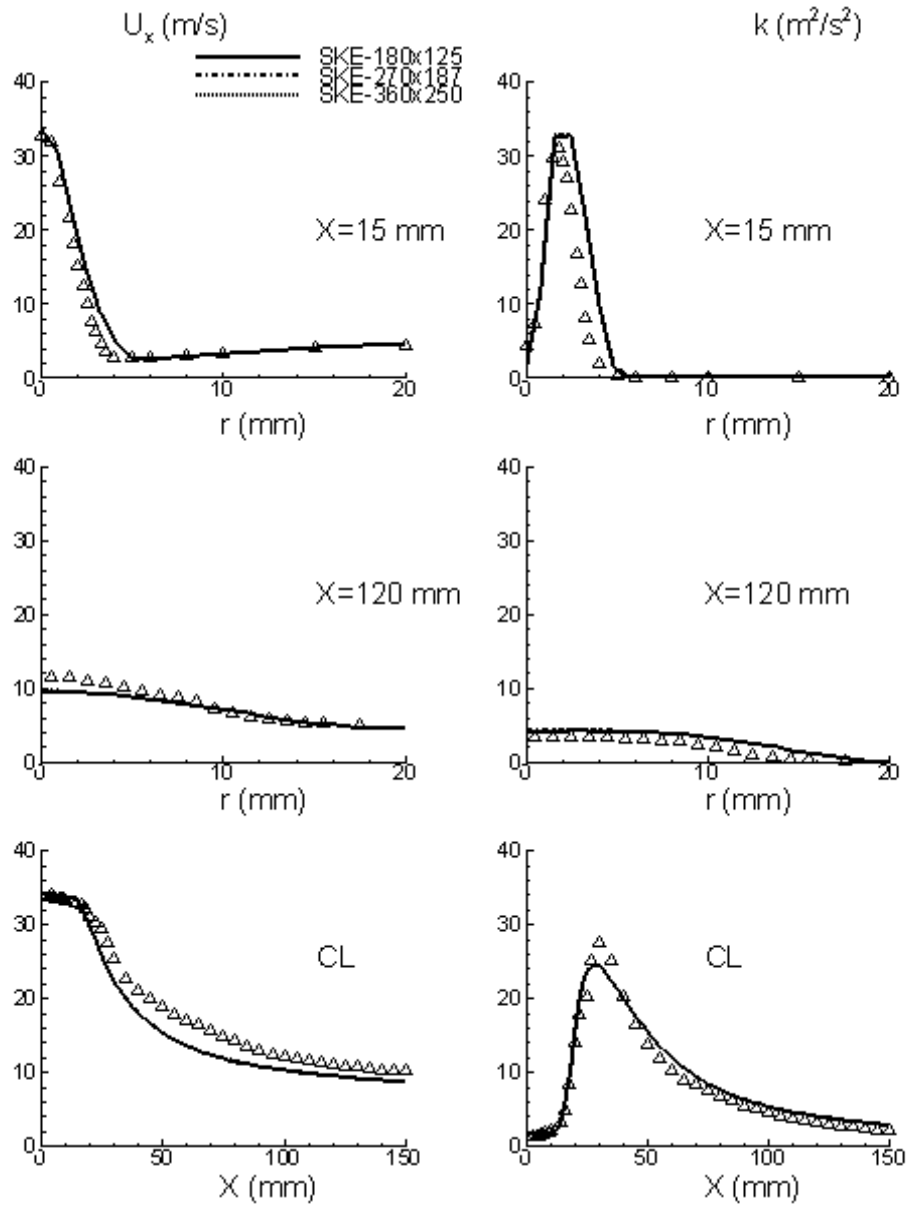

Figure 4. Demonstration of grid independence predictions for mean axial velocity ($U_x$) and turbulent kinetic energy ($k$) of DJHC_I-S at Re=4100. Standard k-ε model with EDC model and Correa chemical mechanism for three different grids. In all figures symbols are measured data and lines are model predictions.

The common modification of the SKE consisting in adjustment of the value of a model constant from $C_{\varepsilon 1} = 1.44$ to $C_{\varepsilon 1} = 1.6$, is found to improve the mean axial velocity prediction, but is leading to an over-prediction of the *k* profile. The SKE and RKE both perform equally well at Re=4100, but at Re=8800 the RKE gives a slightly better prediction for the trends of center line peak value of axial velocity and turbulent kinetic energy (Fig. 5). This is in agreement with the known fact that the RKE performs well compared to SKE in round jet calculations not only for non-reacting turbulent flows [26], but also for reacting turbulent flow calculations [38, 39].

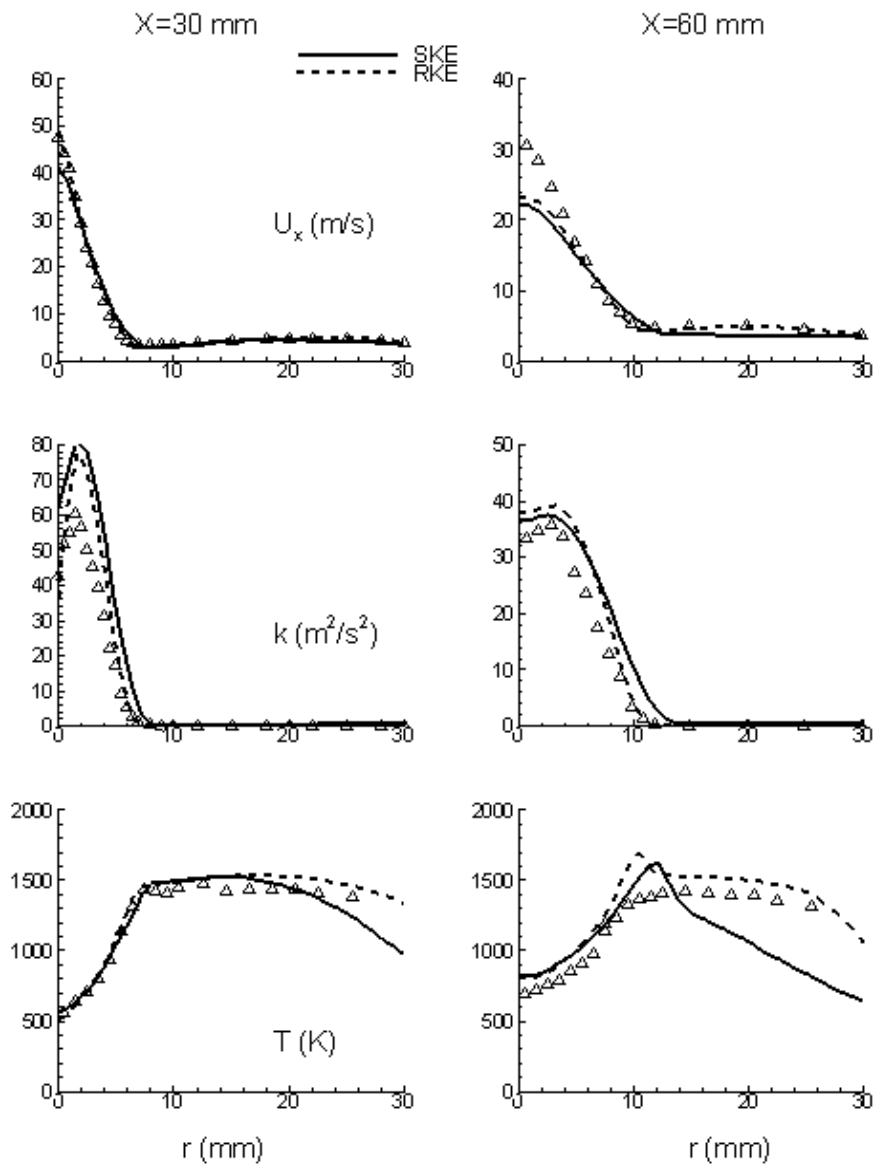

Figure 5. Radial profiles of mean axial velocity ($U_x$), turbulent kinetic energy (*k*) and mean temperature (T) at x=30mm and x=60mm from the jet exit for DJHC_I-S (Re=8800). Comparison of predictions by different turbulent model in combination with the EDC model with the Correa mechanism. Symbols are measurements and lines are predictions.

On the basis of the quality of the predictions of the velocity statistics no decisive advantage of choosing the RKE rather than the SKE appears. Therefore in the next section we also compare their predictions of the mean temperature fields (already shown in Figure 5).

Figure 6 shows the centre-line profiles of mean axial velocity ($U_x$) and turbulent kinetic energy ($k$). The simulation with the Correa and DRM19 mechanisms are done respectively using Fuel I and Fuel II (see Table 2). The difference is unimportant here though. As observed, centerline mean axial velocity profile is well predicted up to x=30 mm and thereafter the center line peak value appears to be slightly under-predicted further downstream, while the overall predictions in the mixing layer are in reasonably good agreement along the axial direction. The under-estimation of axial velocity at the downstream locations is associated with the over-estimation of mean temperature (Fig. 9). As a consequence, mean density value decreases; so that the convection ρU decreases at the axis but at the same time counter-acting diffusion becomes higher. Hence, the dominant diffusive force spreads the jet more rapidly in the radial directions and results in under-prediction of center line peak value. The turbulent kinetic energy profile ($k$) is well predicted up to x=15 mm and thereafter shows some under-predictions along the centerline up to x=40 mm; however it matches well further downstream.

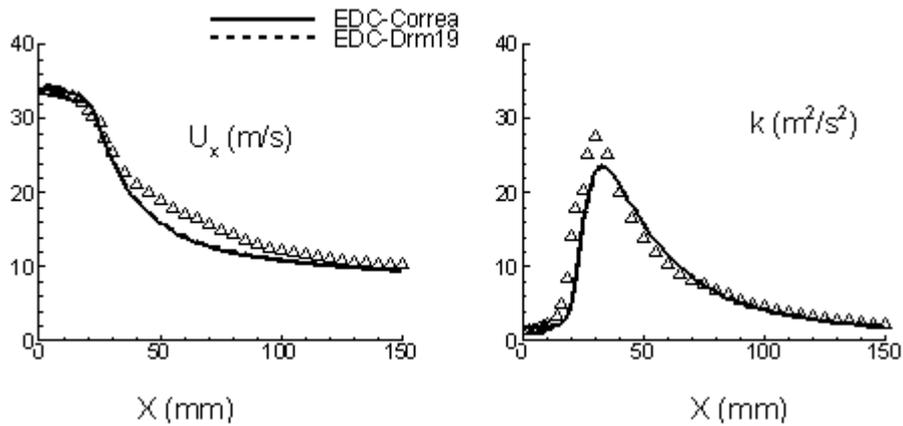

Figure 6. Centre-line profiles of mean axial velocity ($U_x$) and turbulent kinetic energy ($k$) for DJHC_I-S (Re=4100). Comparison of predictions using the RKE model and the EDC model with different chemical mechanisms. Symbols are measurements and lines are predictions.

An important quantity in the description of the mixing of momentum between fuel jet and coflow is the Reynolds stress. Figures 7 and 8 depict the Reynolds stress components from the Boussinesq hypothesis (Eq. 3), obtained by post-processing the computational results, and compared with measurements. We have included two sets of measured data obtained from a complete traverse of the flame. At both sides of the nominal symmetry axis slightly different experimental profiles are found.

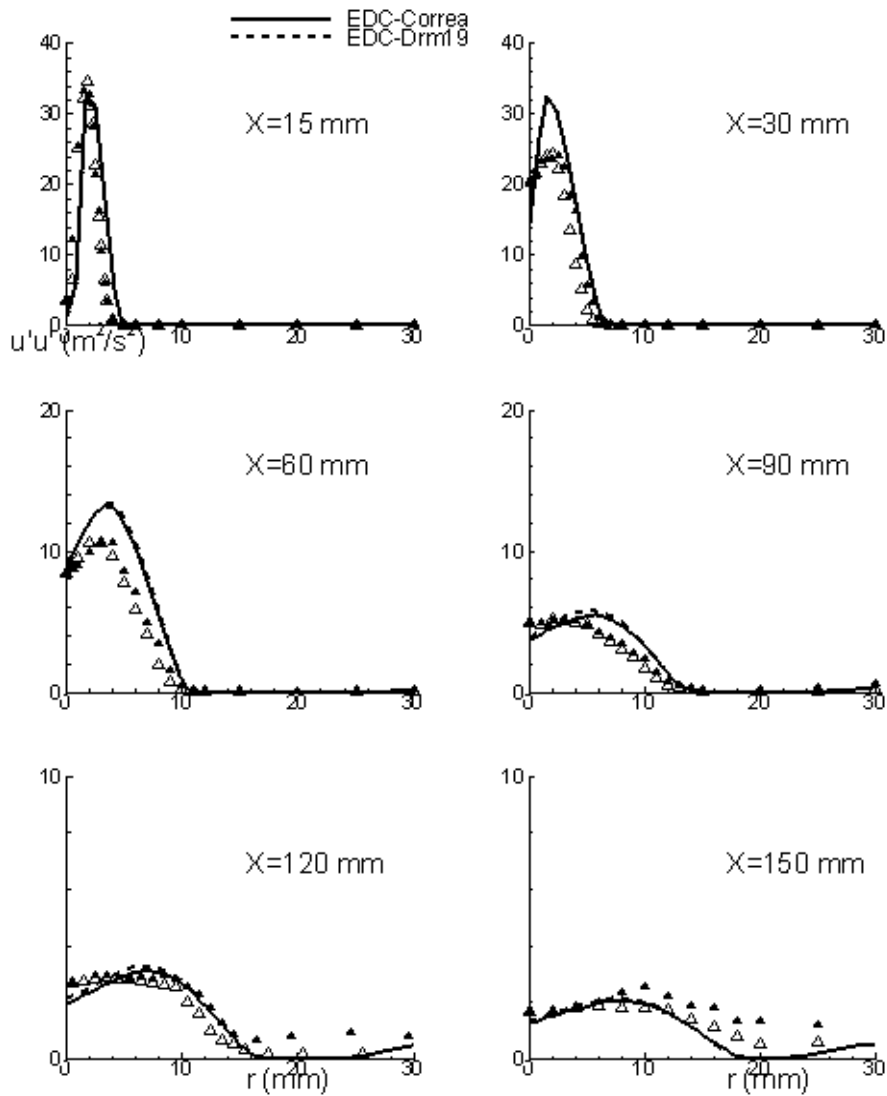

Figure 7. Radial profiles and the centerline profile of Reynolds stress components ($u'u'$) for DJHC_I-S (Re=4100). Predictions using the RKE model and the EDC model with two different chemical mechanisms. Symbols ($\Delta$: $0 \leq r \leq 30$, ▲: $-30 \leq r \leq 0$) are measurements and lines are predictions.

As observed, normal stress components ($u''u''$) are well predicted at x=15 mm and further downstream (x=120 and 150 mm), but show over-prediction at the other locations (x=30, 60 and 90 mm), especially in the shear layer (r<20 mm). This error can be considered typical for *k-ε* models. The Reynolds shear stress ($u"v"$) is in good agreement with measurements as shown in Fig. 8 which clearly shows the location of the shear layer between the fast fuel jet and the slow coflow.

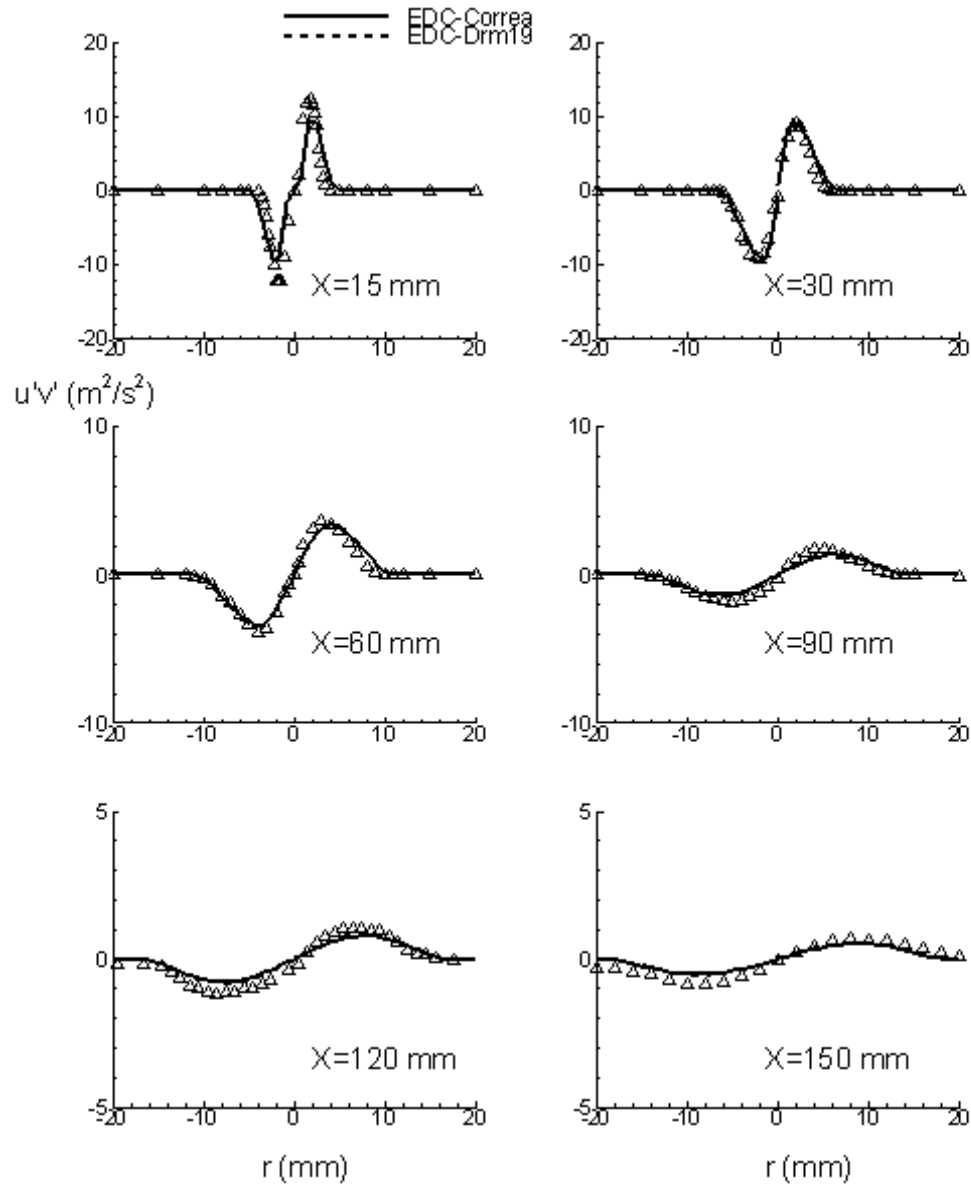

Figure 8. Radial profiles and the centerline profile of Reynolds stress components (*u'v'*) for DJHC_I-S (Re=4100). Predictions using the RKE model and the EDC model with different chemical mechanisms. Symbols are measurements and lines are predictions.

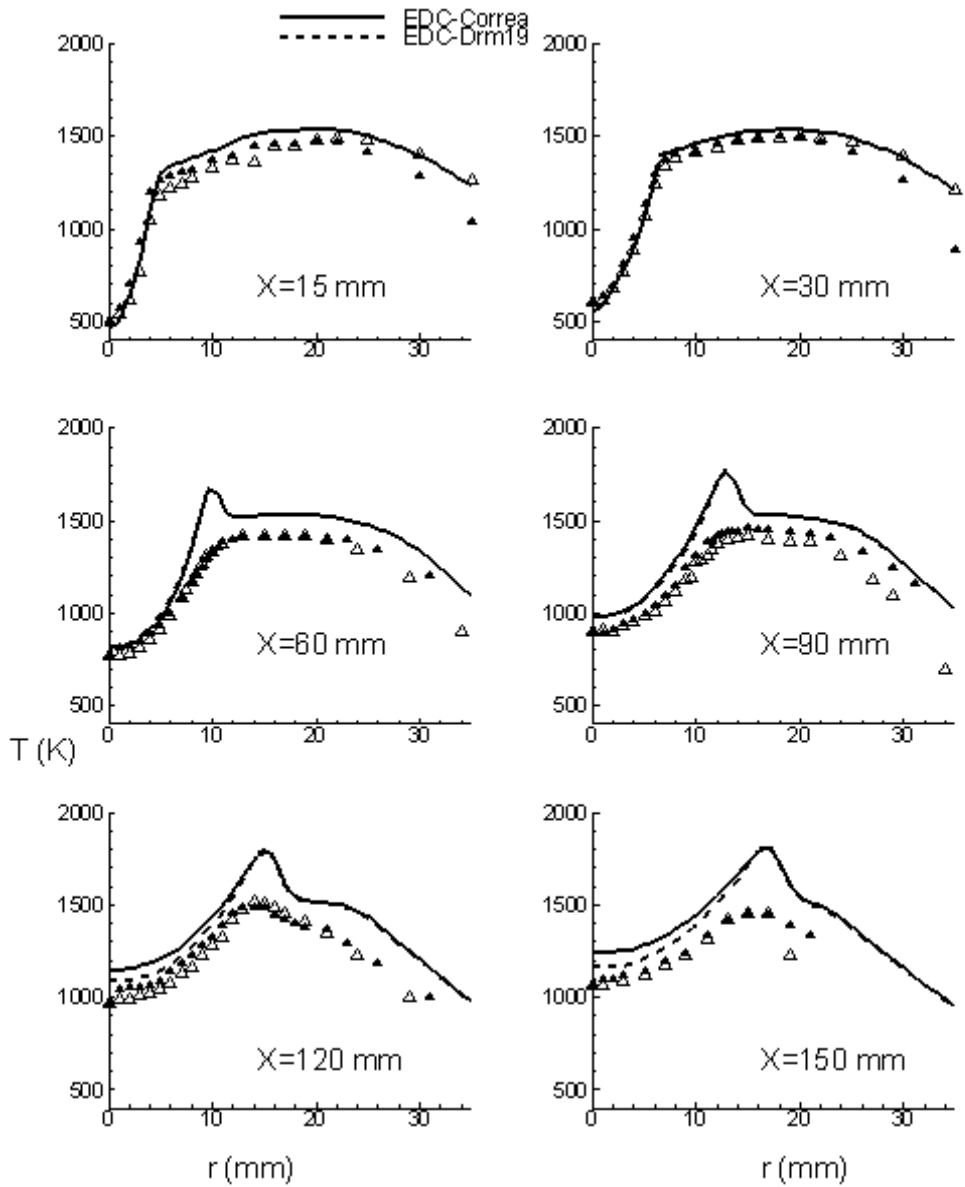

Figure 9. Radial profiles of mean temperature (T) for DJHC_I-S (Re=4100). Comparison of predictions using the RKE model and the EDC model with two different chemical mechanisms. Symbols (Δ:0≤r≤35, ▲: -35≤r≤0) are measurements and lines are predictions.

## 4.3 Prediction of temperature statistics

Substantial differences between predictions by different turbulence models and between models and experiments are found in mean temperature profiles, as shown in Fig. 5. Two problems appear. Firstly, both SKE and RKE predict a peak at about r=12 mm in the radial profile of mean temperature at x=60 mm. The temperature peak in the radial profile at 60 mm is a sign of too early ignition in the simulations and is discussed in more details below. Secondly, the SKE under-predicts mean temperature at large values of r (r>25 mm for x=20mm, and r>16 mm for x=60 mm). The explanation is that the SKE is giving an over-prediction of entrainment bringing in too much cold air from the surroundings. In fact, in the experimental study [7, 8] a decrease in lift-off height with increase in Reynolds number is observed, and this is explained as a consequence of entrainment of surrounding coflow by the fuel jet in the case of a radial mean temperature gradient in the coflow. The better prediction of radial profile of mean temperature in the outer region is a decisive advantage of the RKE over the other considered turbulence models and the RKE-results will be used in the subsequent analysis.

Since the mixing of momentum between jet and coflow appears to be well described by the RKE turbulence model, also the mixing of specific enthalpy can be expected to be well described. Possible deviations between measurements and predictions in radial profiles of mean temperature then are not caused by the model for mixing of jet and coflow, but are related to the description of the combustion process. Figure 9 shows the radial profiles of mean temperature at different axial locations. The asymmetry in the measurements in the outer region (r>25 mm) needs to be noted while comparing the predictions with data, especially up to x=90 mm. As mentioned before, ignition is predicted too early, as is clear from the temperature peak at the outer edge of the shear layer in the stoichiometric zone.

The center line mean temperature is in better agreement with experiment in the simulation with fuel containing $C_2H_6$ as shown in Fig. 9 at x=120 mm and x=150 mm. Differences between the predictions by the two different chemical mechanisms are also visible in the downstream region. Combining information from Figs. 7 and Fig. 8, it is clear that the momentum shear layer and the main reaction zone are at different radial position. The temperature peaks appear in the zone where fuel and coflow are present in stoichiometric proportion (Fig. 9).

The mean temperature is well predicted at x=15 and x=30mm, but it tends to be over-predicted along the inner edge of the shear layer at the downstream locations including peaks in radial temperature profiles (10<r<20, Fig. 9). This is due to the overestimation of mean reaction rate in the EDC model, sign of a too early ignition. It is noteworthy to be remarked that the consistent over-prediction in the temperature profiles in the outer edge of shear layer (r>15 mm) is not due to any radiative effects in the flame zone. It has also been checked by calculations with P-1 radiation model as well as discrete ordinates method [29] in conjunction with weighted-sum-of-gray-gases model that the maximum temperature difference between calculations with and without radiation taken into account is about 50K.

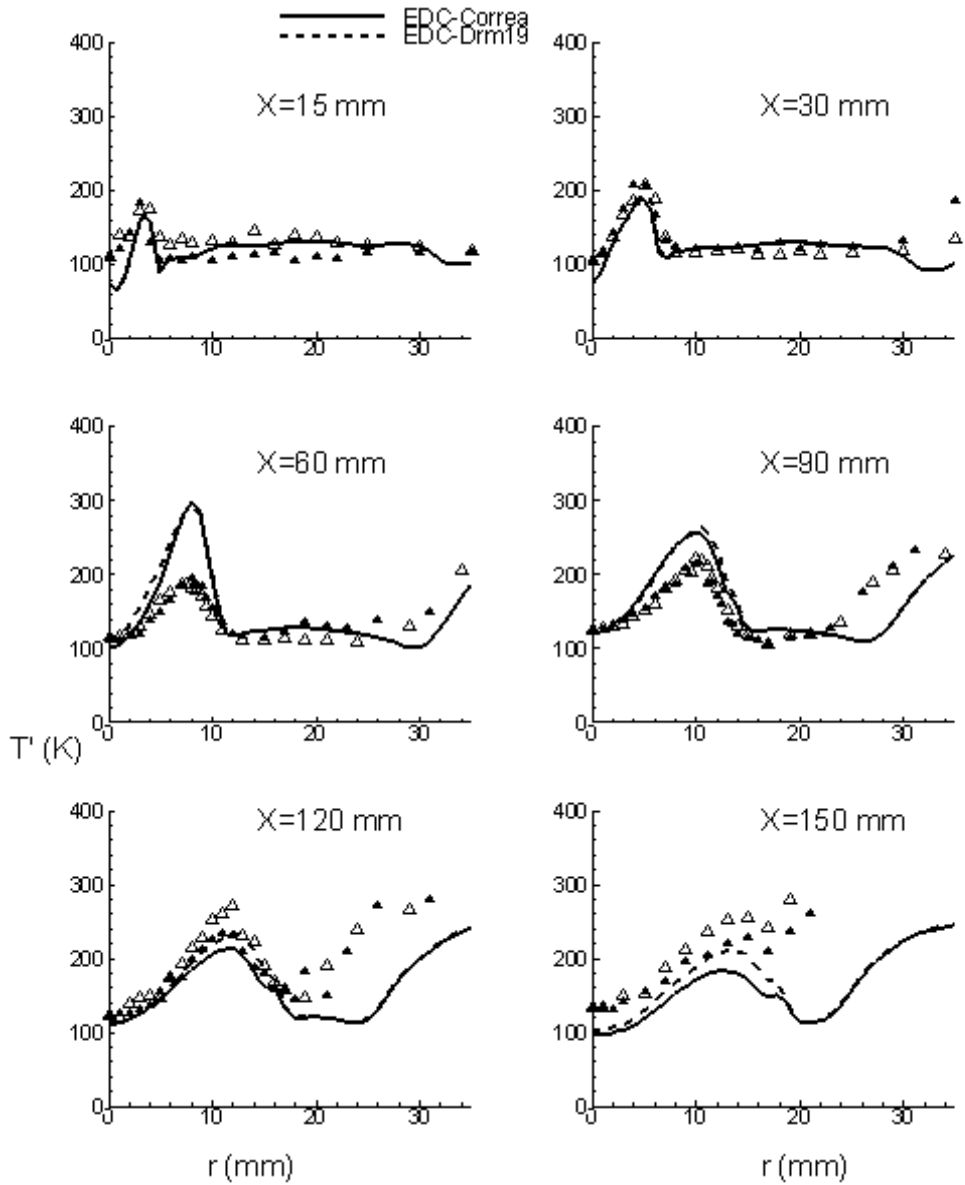

Figure 10. Radial profiles of the standard deviation of temperature (T') for DJHC_I-S (Re=4100). Comparison of predictions using the RKE model and the EDC model with two different chemical mechanisms. Symbols (Δ:0≤r≤35, ▲: -35≤r≤0) are measurements and lines are predictions.

Figure 10 shows the standard deviation of temperature at different axial locations. The predictions are in good agreement with the experimental data at x=15 mm and x=30 mm, but there is a general trend of over-predictions at distances 60 mm and 90 mm. These over-predictions can be associated with the balance maintained between production and destruction of temperature fluctuations as characterized by T' (Eq. 14). The production of higher variance (Fig. 10, x≥60mm) is caused by higher gradients of mean temperature (Eq. 14). On the other hand, the over-prediction of turbulent kinetic energy in the shear layer (higher eddy viscosity) leads to a higher diffusion of the variance (Eq. 14) at the same locations (x≥60mm) and also contributes to the production. However, at the same time the third term in Eq. (14) acts as a destruction of variance. Therefore, this combined effect of production and destruction terms (Eq. 14) causes the over-prediction of standard deviation at further downstream locations. Hence, in the present formulation the proper prediction of the standard deviation of temperature largely depends on the predicted mean temperature and the turbulence model, and not directly on reaction rates. This explains why different chemical mechanisms do not induce substantial differences in predictions of the temperature RMS-value.

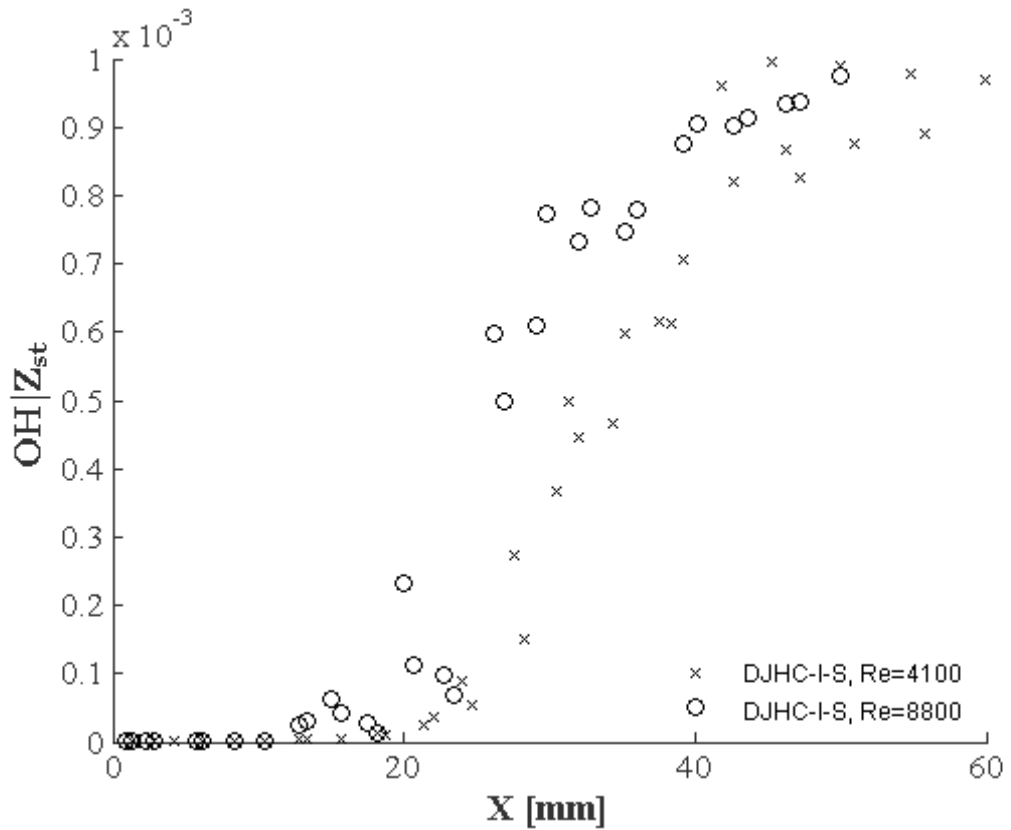

Figure 11. Profile of the mass fraction of OH, in the computational cells where $0.065 < \tilde{Z} < 0.075$, as function of height above the fuel nozzle exit. Model predictions using turbulence model RKE and EDC model with kinetic mechanism DRM19.

## 4.4 Prediction of trends in lift-off height

In the experimental study of the DJHC flames, Oldenhof et al. have found that the height at which the first ignition kernels appear is decreasing with increasing jet Reynolds number [7, 8]. As an explanation for this trend the increase of entrainment with increase of jet velocity, coupled with the positive radial temperature gradient at X=0 were brought forward. Since the flow field including entrainment is predicted quite well using the RKE model in combination with the EDC model it can be expected that this mechanism is correctly described. In a turbulent flame there is a fluctuating contour of stoichiometric mixture composition. Because using the EDC model only mean mass fractions are known, only an approximate mean stoichiometric contour can be identified, namely by applying Bilger's formula [40] for mixture fraction, taking fuel and air as base streams, but using mean mass fraction as input and looking at the contour where this approximate mean mixture fraction $\tilde{Z}$ is near stoichiometric. The contour obtained in this way adequately characterizes the ignition region for our purpose here. As a marker for lift-off height we use the mean OH mass fraction. Figure 11 shows the profile of $\tilde{Y}_{OH}$ versus height above the fuel nozzle exit for computational cells in the region where $0.065 < \tilde{Z} < 0.075$. This results shows that the EDC model correctly predicts the decreasing trend of lift-off height with increase in fuel jet velocity.

A critical assumption in the explanation given in Ref. [8] of trends in lift-off height as a result of entrainment effects is that temperature hardly changes along the streamlines in the coflow region. This assumption is corroborated in the context of the present RANS simulations. Figure 12 shows the profiles of mean temperature and mean mixture fraction versus stream function $\Psi(r)$. As in Ref. [8] this stream function is the mass flow rate originating from the coflow integrated up to a radial distance $r$. By definition it is zero at X=0, $r$=2.5mm. It can be seen that in the coflow region the mean temperature at a certain value of stream function is independent of axial distance X. The increase of temperature after ignition appears at the values of stream function where the mixture fraction is near stoichiometric.

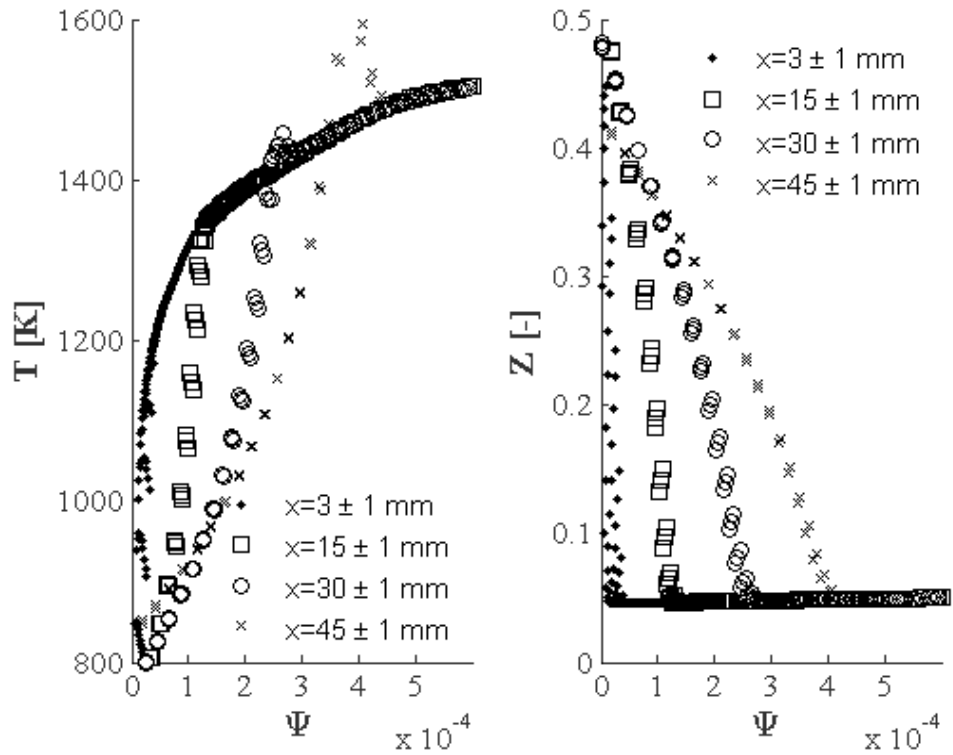

Figure 12. Profiles of mean temperature versus mean stream function (left) and mean mixture fraction versus mean stream function (right) at different axial locations. Model predictions using turbulence model RKE and EDC model with kinetic mechanism DRM19.

## 4.5 Sensitivity to EDC model constants

The analysis of section 3 has shown that the predictions of the Fluent EDC model may be sensitive to the value of the model constants in a situation with low turbulent Reynolds number. Here we first show that in the case under investigation the model predictions indeed are sensitive to the value of model parameters and then proceed to further analysis.

A change of either of the two EDC model constants is found to lead to better agreement in the predictions of mean temperature profiles. In the first case, the time scale constant is increased from default value of $C_\tau = 0.4082$ to $C_\tau = 3.0$. In the second case the volume fraction constant is decreased from default value of $C_\xi = 2.1377$ to $C_\xi = 1.0$. Figure 13 depicts the mean temperature and RMS-value of temperature profiles using the different EDC model parameters. It is clear that using modified EDC parameters substantially improve the results. Importantly, the change of value of model constants does not visibly affect the profiles of mean axial velocity and turbulent kinetic energy (not shown in figures) profiles.

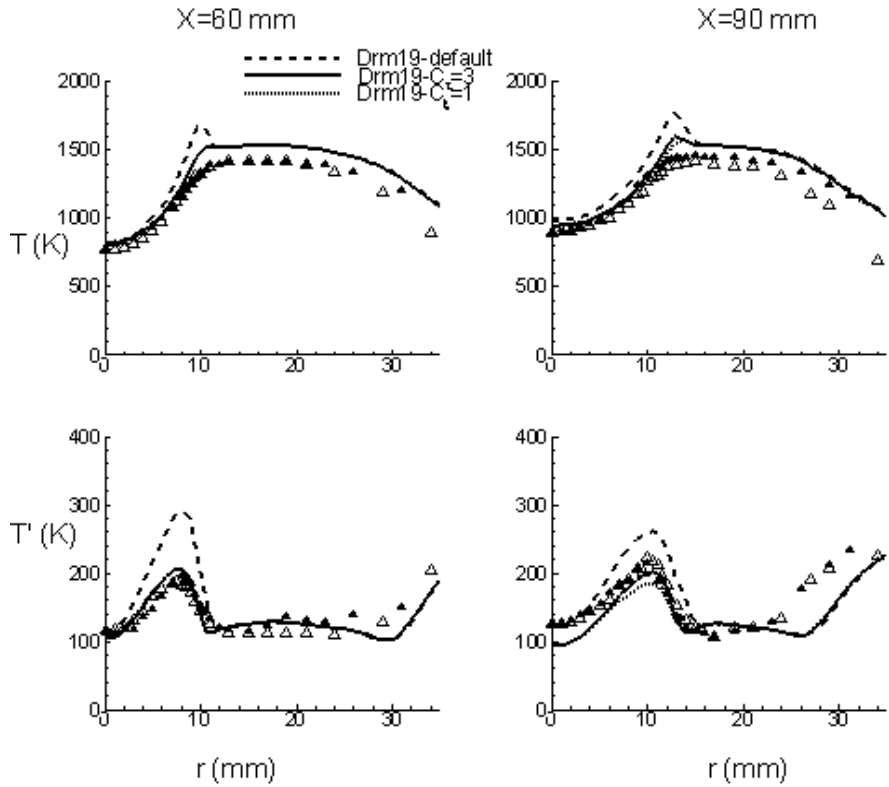

Figure 13. Radial profiles of mean Temperature (T) and RMS of temperature (T') for DJHC_I-S (Re=4100). Comparison of results using different values of EDC model parameters (with turbulence model RKE and kinetic mechanism DRM19). Symbols (Δ:0≤r≤35, ▲: -35≤r≤0) are measurements and lines are predictions.

In order to investigate whether a change of model constants has qualitatively the same effect in other cases, characterized by different oxygen content and temperature of the coflow we now also consider the flame DJHC_X-S at Re=4600. For this flame configuration, the $O_2$ concentration in the coflow is higher (10.9 %) and the maximum coflow temperature is lower (1395 K) compared to case DJHC_I-S (See Table 1). The general trends of the predictions with default or modified values of the EDC model constants are found to be similar to those found in DJHC_I-S, apart from lower temperatures in the flow field due to lower coflow temperature. Figure 14 shows the Mean temperature (T) profiles at different axial locations. The EDC model with default values of model constants appears to over-predict the mean temperature including the occurrence of early ignition. The predictions of the EDC model with $C_\tau = 0.4082$ modified to $C_\tau = 3.0$ shows better agreement. At x=90 mm the too early ignition is avoided. At larger axial distance an overall over-prediction of mean temperature remains.

It can be concluded that the considered change of model constant is adequate to have ignition at the right axial distance for both DJHC_I-S and DJHC_X-S, but not to reach globally good results. The question arises whether this change in value of model constants has physical meaning related to a change of combustion regime in MILD combustion, compared to standard combustion. Before concluding anything on this, it is worth investigating the role of turbulent Reynolds number in the present burner configuration.

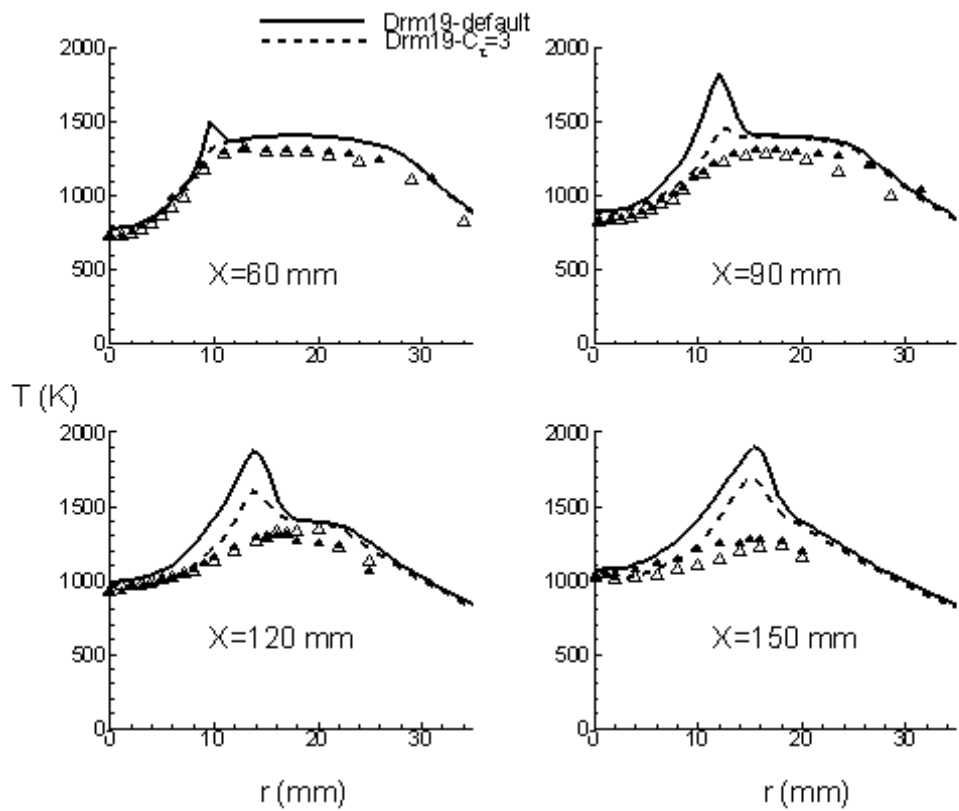

Figure 14. Radial profiles of mean Temperature (T) for DJHC_X-S (Re=4600). Comparison of results using different values of EDC model parameters (with turbulence model RKE and kinetic mechanism DRM19). Symbols (Δ:0≤r≤35, ▲: -35≤r≤0) are measurements and lines are predictions.

### 4.5 Effect of low turbulent Reynolds number

#### 4.5.1 Analysis of the mean reaction rate

The analysis of the expression of the mean reaction rate of the EDC model in section 3.3.2 has clearly shown that in a situation with low turbulent Reynolds number it is sensitive to the value of the model constants. To analyze whether this effect plays a role in the predictions of mean temperature shown in Figs. 13 & 14, a few additional plots are shown in the Fig. 15.

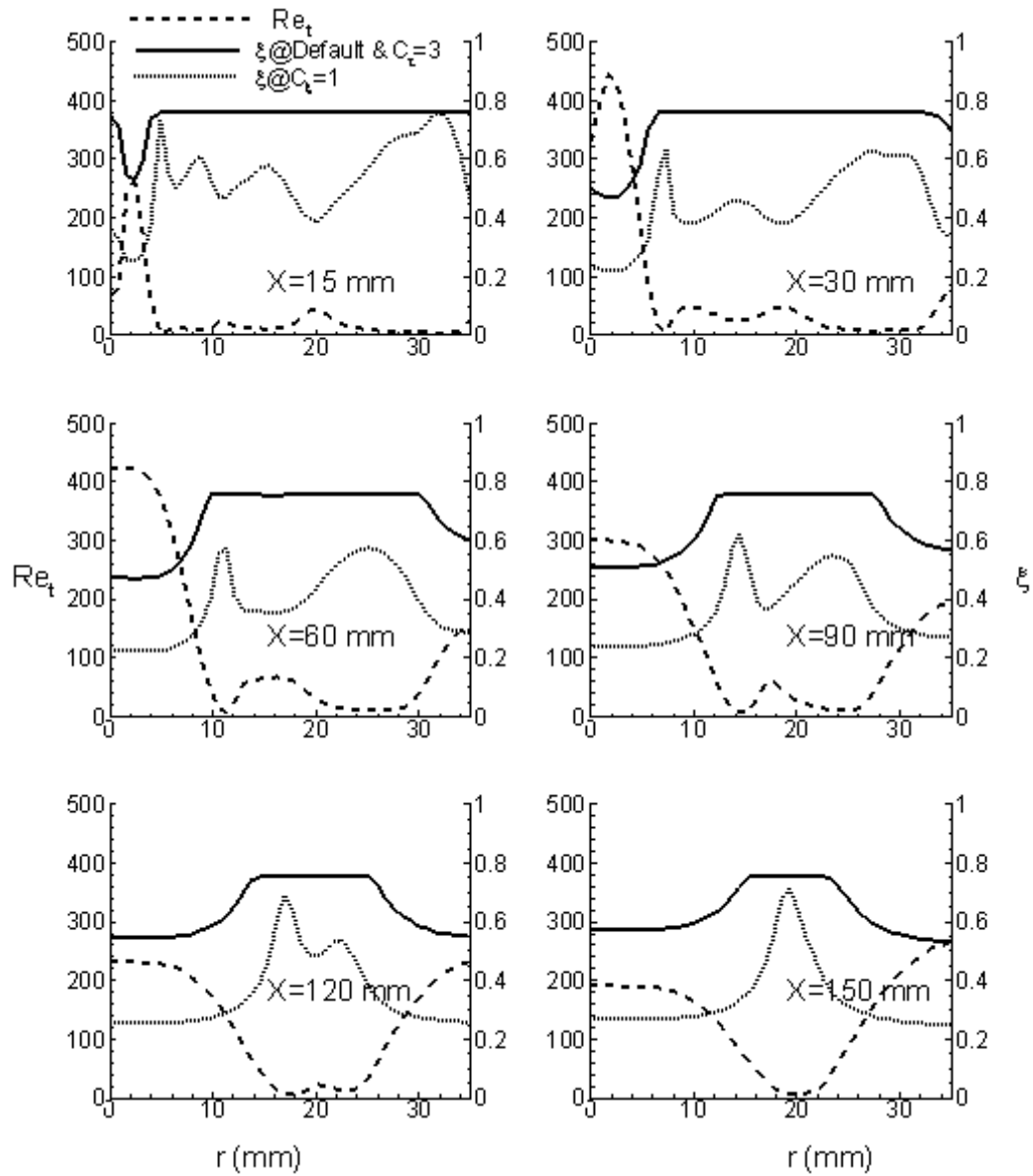

Figure 15. Radial profiles of $Re_t$ (left vertical axis), $\xi$ (Eq. 8) (right vertical axis) for DJHC_I-S (Re=4100). Comparison of results using different values of EDC model parameters (with turbulence model RKE and kinetic mechanism DRM19).

Figure 15 shows the value of turbulent Reynolds number ($Re_t$) and the value of the EDC parameter $\xi$ (Eq. 8) as function of radial position for different values of model constants $C_\tau$ and $C_\xi$. The value of $Re_t$ is not visibly depending on the values of the model constant and is represented by a single curve. This is quite obvious since $Re_t$ contains the information of the turbulent flow field that is only very weakly influenced by the settings of the model constants in the EDC model. As can be seen in the figure, $Re_t$ takes values lower than the critical value $Re_t$~65 (see section 3.3.2) in part of the coflow region. At default value of the EDC model constants the unphysical behavior of the mean reaction rate is prevented by clipping the EDC model parameter ξ (Eq. 8) such that R=1, i.e. for any value of $Re_t$ lower than 65, ξ has the value 0.75. In the core of the jet and the shear layer the Reynolds number is sufficiently high to stay away from the sensitive region (Fig. 8). The main reason for the low values of turbulent Reynolds number is the low turbulence and high temperature of the coflow. In the reaction zone this is enhanced by the increase of viscosity with increase in temperature. As a consequence, the radial region where clipping occurs for x≥30 mm extends outwards from the radial location of the peak in mean temperature profiles and contains the location of the reaction zone.

When the time scale constant $C_\tau$ is increased from default value of $C_\tau = 0.4082$ to $C_\tau = 3.0$, the profile of ξ as function of radial position does not visibly change. But this change in model constant lowers the mean temperature peaks due to lower reaction rates (Eq. 7). On the other hand when the value of the model constant $C_\xi$ is reduced, the radial profile of ξ drastically changes and ξ is never reaching the value where clipping is needed (Fig. 15). This leads to better agreement in mean temperature profiles than the default value as shown in Fig. 13.

From this analysis we conclude that the application of the EDC model to the current jet-in-hot-coflow problem is showing new features of the model, not visible in the application to high Reynolds number flows. The analysis given in section 3.3.2 shows that in regions in the flow with low Reynolds number, where large and small scales of turbulence come very close to each other, the original motivation to introduce the model is no longer valid, i.e. there no longer is a clear separation between large and small scales in turbulence, with reaction structures of the size of the small scales. As explained in section 3.3.2, in those regions of the flow the EDC reaction rate has to be interpreted as a smoothed laminar reaction rate, and the proper evaluation of its performance should be done comparing the Kolmogorov scale to the chemical time scales. Good performance can be expected only for those phenomena which are dominated by the chemical time scales slower than the Kolmogorov scale. This analysis applies to the DJHC burner studied here, but it may also apply to the calculations of the experiments by Dally et al using EDC reported in the literature [10-11, 13-14].

**4.5.2 Analysis of the effects of laminar viscosity and diffusivity**

At low turbulent Reynolds number the laminar viscosity is not much smaller than the turbulent viscosity and the model used to represent it must be accurate, in particular in combustion its temperature dependence plays a role. Another molecular effect in the simulations can come from the species diffusivities. In order to analyze the sensitivity of the predictions to the temperature dependence of viscosity and the model for species diffusivities, additional simulations were carried out using the RKE turbulence model with DRM19 chemical mechanism with default model parameters. The results are briefly summarized here.

Comparison of results using different models for viscosity (Sutherland's law and constant viscosity $1.72 \times 10^{-5}$ $kg\ m^{-1}s^{-1}$ show that the temperature dependent viscosity had little impact on the flow field and there are no substantial differences observed between the two models. Some notable differences are observed in turbulent Reynolds number (Eq. 5) and EDC model constant $\xi$ (Eq. 8) predictions using temperature dependent viscosity though. The combined effect of viscosity on $Re_t$ (Eq. 5) and $\xi$ (Eq. 8) is such that the mean reaction rate is not strongly affected and this explains why no substantial differences are observed in flow field predictions.

In the preceding sections we have presented results using constant species diffusivities (constant dilute approximation). But we have checked the sensitivity to the species diffusivity model by comparing with a calculation using multi-component diffusion and found that the differences are even smaller than the difference between simulations using constant or temperature dependent viscosity and therefore our conclusions on the accuracy of the turbulence and chemistry models are independent of the choice of diffusivity model. The situation may be different in the case of fuel with a substantial amount of hydrogen in the fuel, as pointed out by Dally et al. [4] and studied also by Mardani et al. [14].

## 5. Conclusions

In this work results of numerical investigations of the Delft-Jet-in-Hot-Coflow (DJHC) burner emulating MILD combustion behavior have been reported. Two different flame conditions (DJHC_I-S and DJHC_X-S) corresponding to two different oxygen levels (7.6% and 10.9% by mass) in the hot coflow have been simulated. The case of coflow with 7.6% oxygen, is computed for two different jet Reynolds number (Re=4100 and Re=8800). In all cases comparison with experimental data by Oldenhof et al [7-8] is made. A detailed study on the performance of the EDC model in combination with two-equation turbulence models and chemical kinetic schemes for about 20 species (Correa mechanism and DRM19 mechanism) has been carried out. The following conclusions can be inferred from the present investigation:

The predicted mean velocity and turbulent kinetic energy fields are in good agreement with the measurements. As expected for this round jet system the realizable k-ε model performs better than the standard k-ε model. This is more clear for the jet with Re=8800 than the jet with Re=4100. The most clear advantage of using the realizable k-ε model is the better prediction of the radial mean temperature profile for the Re=8800 case. In the outer region of the coflow entrainment of cold air from the surroundings plays a role and this is better predicted by the realizable k-ε (RKE) model. The renormalization group k-ε (RNG) model does not perform well.

The predicted mean temperature field shows systematic deviations from experimental results. The radial mean temperature profiles show a peak due to too early ignition. In other words the EDC under-predicts lift-off height. However the decreasing trend of lift-off height with increase of jet exit velocity is correctly predicted. The overall temperature level is somewhat over-predicted in the downstream region. The prediction of too early ignition can be avoided by using modified values of the EDC model parameters: either volume fraction constant $C_\xi$ or time scale constant $C_\tau$. The quality of the prediction of standard deviation of temperature by using an additional model equation is mainly controlled by the quality of the prediction of the mean temperature, and not by the modeling of the reaction rate.

A theoretical analysis of the reaction rate expression of the EDC model has shown that the derivation from arguments assuming the presence of a turbulent spectrum no longer holds at values of turbulent Reynolds number lower than 65 (for the default values of model constants). For lower $Re_t$ the general form of the EDC reaction rate has to be clipped and the meaning of the clipped reaction rate is that of a smoothed laminar reaction rate. Model predictions can become sensitive to this procedure. This evidently plays a critical role in the too early prediction of ignition in the DJHC burner. In the experiments simulated here, with natural gas as fuel, the effects of differential diffusion are found to be small.

The observed properties of the EDC rate expression in regions of the flow where the turbulent Reynolds number is low, could also play a role in EDC model simulations reported in [10-11,13-14] of the experiments by Dally et al [4].

**Acknowledgements**

We gratefully acknowledge financial support of this project by the Dutch Technology Foundation (STW) (project 06910).

**Appendix**

In this appendix more information is given on the burner design.

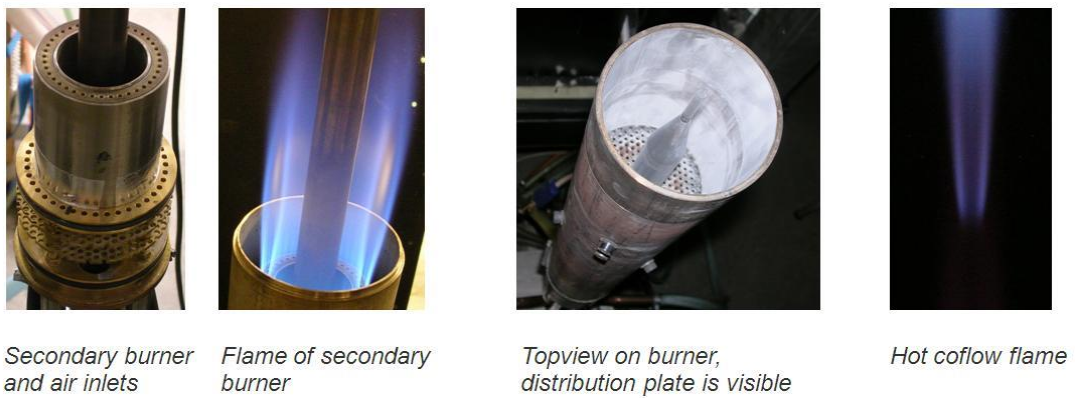

Figure 16: Photographs illustrating the burner design and the flame structure

Figure 16 shows photographs illustrating the construction of the burner and the resulting flames. The first picture from the left shows the internals of the secondary burner: a ring burner for injecting a premixed fuel/air mixture and at larger radius and further upstream a ring of nozzles for additional air injection. The appearance of the flame of the secondary burner when it is not enclosed is shown in the second photo from the left. When the burner is fully installed it impinges on the distribution plate, visible in the third photograph. The distribution plate promotes heat loss (by radiation from the heated plate) and mixing of combustion products and air in the coflow, leading to the overall properties listed in Table 1. An important feature of this configuration is that the radial profile of mean temperature and oxygen concentration at the height of the main burner nozzle is not uniform, which has to be taken into account in simulations.